\documentclass[twocolumn]{article}

\usepackage{amsmath, fullpage}
\usepackage{authblk}
\usepackage{graphicx}
\usepackage{comment}
\usepackage{amsmath}
\usepackage{amssymb}
\usepackage[font=small, labelfont=bf]{caption}
\usepackage{subcaption}
\usepackage[dvipsnames]{xcolor}
\usepackage[normalem]{ulem}
\usepackage{dblfloatfix} % put in the preamble
\usepackage{abstract}
\usepackage[version=4]{mhchem} % for chemistry formatting if needed
\usepackage{achemso}

\title{Enhanced Secondary-Electron Detection of Single Ion Implants in Silicon Through Thin SiO$_2$ Layers}
\author[1]{E. B. Schneider}
\author[1]{O. G. Lloyd-Willard}
\author[2]{K. Stockbridge}
\author[1]{M. Ludlow}
\author[1]{S. Eserin}
\author[1]{L. Antwis}
\author[1]{D. C. Cox}
\author[1]{R. P. Webb}
\author[1]{B. N. Murdin}
\author[1]{S. K. Clowes}

\affil[1]{Surrey Ion Beam Centre, University of Surrey, Guildford GU2 7XH, UK}
\affil[2]{Ionoptika Ltd., Unit B6, Millbrook Cl, Chandler's Ford, Eastleigh SO53 4BZ, UK}

\begin{document}

%\maketitle

\twocolumn[
\maketitle
\begin{onecolabstract}

Deterministic placement of single dopants is essential for scalable quantum devices based on group-V donors in silicon. 
We demonstrate a non-destructive, high-efficiency method for detecting individual ion implantation events using secondary electrons (SEs) in a focused ion beam (FIB) system. 
Using low-energy Sb ions implanted into undoped silicon, we achieve up to 98\% single-ion detection efficiency, verified by calibrated ion-current measurements before and after implantation. 
The technique attains $\sim$30 nm spatial resolution without requiring electrical contacts or device fabrication, in contrast to ion-beam-induced-current (IBIC) methods. 
We find that introducing a controlled SiO$_2$ capping layer significantly enhances SE yield, consistent with an increased electron mean free path in the oxide, while maintaining high probability of successful ion deposition in the underlying substrate. The yield appears to scale with ion velocity, so higher projectile mass (e.g. Yb, Bi etc) requires increased energy to maintain detection efficiency.
Our approach provides a robust and scalable route to precise donor placement and extends deterministic implantation strategies to a broad range of material systems and quantum device architectures.

\end{onecolabstract}
\vspace{1em}
]

%\linenumber

%\begin{abstract}
%\end{abstract}

\section{Introduction} 

The ability to position and activate individual atoms within a solid with atomic precision represents a defining goal of nanotechnology \cite{Schofield2025}. 
Such control underpins a new generation of quantum devices in which a single dopant or defect serves as the functional unit. 
Examples include donor-based spin qubits in silicon \cite{Morello2020,chang2024,Yu2025}, single-electron transistors, and single-photon emitters in wide-bandgap hosts. 
Among available fabrication techniques, ion implantation stands out for its ability to introduce dopants with well-defined species, energy, and spatial localization, and is already the cornerstone of conventional semiconductor device processing.

For quantum technologies, however, the regime of interest is that of \emph{single-ion} implantation, where precise control over both position and number is essential. 
Ion emission from most focused ion beam (FIB) or accelerator sources is inherently stochastic and can be described by a Poisson process. 
In this case, the probability of implanting exactly one ion per pulse is limited to 37\%, even under ideal tuning conditions. 
To achieve deterministic placement, each implantation event must therefore be detected in real time, allowing the beam to be terminated once a single ion has arrived. 
This requirement has motivated the development of various high-sensitivity ion impact detection methods.

Current approaches rely primarily on ion-beam-induced current (IBIC) or charge-collection measurements~\cite{Jakob2022,Titze2022}, which demand pre-fabricated device structures and electrical contacts, thereby restricting throughput and material flexibility. 
Secondary electron (SE) detection \cite{Cassidy2021,Murdin2021,Adshead2025} provides a non-destructive, contactless alternative, yet conventional SE-based schemes suffer from lower signal-to-noise ratios and consequently lower efficiency than IBIC for heavy ions at low energies. 
High SE efficiency detection has been achieved for some species/host combinations including for Sb in Si and SiO$_2$ \cite{Adshead2025} but error bars are typically high, and some combinations reportedly exceed 100\% efficiency.
Other techniques \cite{Sahin2017,Racke2022,Stopp2022}, though highly precise, often depend on large-scale accelerator infrastructure and have not yet demonstrated both high confidence and high throughput for single-ion implantation.

In this work we demonstrate high-efficiency SE (up to 98\%) with very small uncertainty for detecting individual Sb  ion implantation events in silicon using SE emission in a FIB system. We chose Sb because it is a prominent choice for silicon qubits.  
We identify a robust SE-yield enhancement from ultrathin SiO$_2$ capping layers and determine the oxide thickness that maximizes the overall implant-and-detect success probability and we explain species/energy trends via Lindhard-Scharff electronic stopping together in terms of changes in escape probability and inelastic mean free path. 
By precisely calibrating the ion current before and after implantation, we obtain a quantitative measure of detection efficiency, exceeding 90\%. 
We further show that introducing a controlled SiO$_2$ capping layer enhances SE yield substantially, which we attribute to an increased escape probability and mean free path of low-energy electrons in the oxide relative to silicon. 
The method achieves nanometre-scale spatial resolution and is inherently compatible with a wide range of host materials, including diamond and SiC. 
This technique thus provides a robust and scalable route for deterministic single-ion implantation, paving the way for non-destructive integration of donor-based quantum devices and colour centres across diverse material platforms.

\section{Experiment} 
If we assume the false positive rate (dark counts per second) can be ignored, then the total number of detection events per unit time is
\(  N = \eta L,
\)
where $L$ is the ion flux (ions per second) and $\eta$ is the detection efficiency, i.e. the ratio of true positives detected to the total number of true positives. 
The total number of detected events per pulse is therefore 
\(
    \nu = N t = \eta L t,
\)
where $t$ is the pulse duration. 
Assuming each ion detection event is independent, the detected events follow a Poisson distribution with mean $\nu$. 
The probability of an apparently empty pulse is therefore given by~\cite{Cassidy2021}
\begin{equation}
    p_0 = e^{-\nu} = e^{-\eta L t}.
    \label{eqn:p0def}
\end{equation}
This expression implies that a graph of $\nu=-\ln(p_0)$ against $\lambda = Lt$ (the number of ions per pulse) yields a slope equal to $\eta$.
The experiment was designed to obtain $p_0(L,t)$.

High-resistivity silicon samples were prepared with SiO$_2$ layers ranging from 2 to 10.4 nm in thickness, deposited by atomic layer deposition (ALD) using an Ultratech/Cambridge Nanotech Savannah S100 system. The ALD process enables highly uniform oxide growth with atomic-layer precision. Deposition was performed on top of the native oxide, and the total SiO$_2$ thickness was measured by spectroscopic ellipsometry, which provided sub-nanometre accuracy in determining the combined oxide thickness.

Ion implantation was carried out using the Single Ion Multi-Species Positioning at Low Energy (SIMPLE) tool (Ionoptika QOne) at the Surrey Ion Beam Centre, with Sb ions implanted at 25keV and 50keV, respectively. These experiments were designed to evaluate single-ion detection efficiency and the yields of secondary particles produced during impact.

\begin{figure*}[!b]
    \centering
    % Left image
    \includegraphics[width=0.49\textwidth]{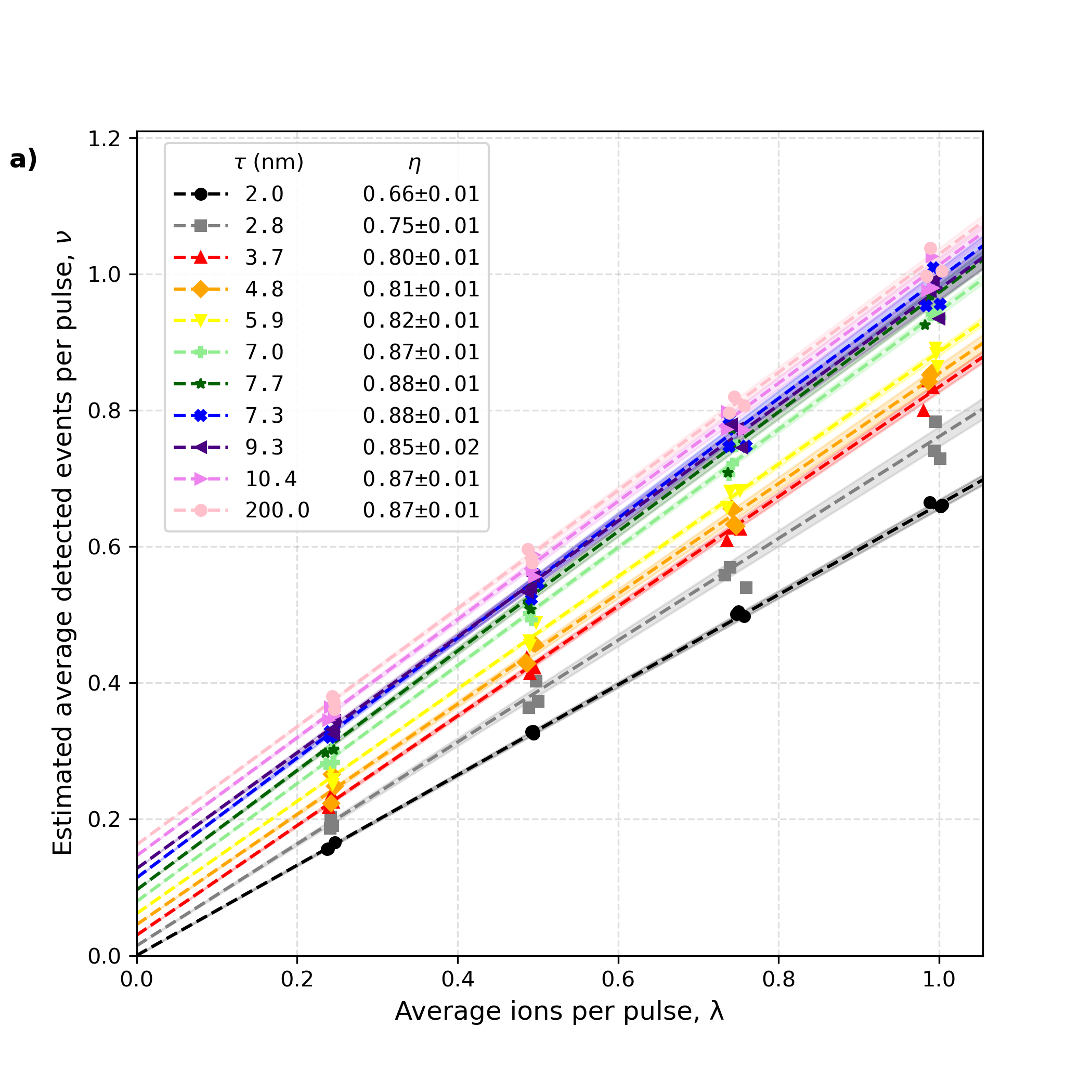}
    \hfill
    % Right image
    \includegraphics[width=0.49\textwidth]{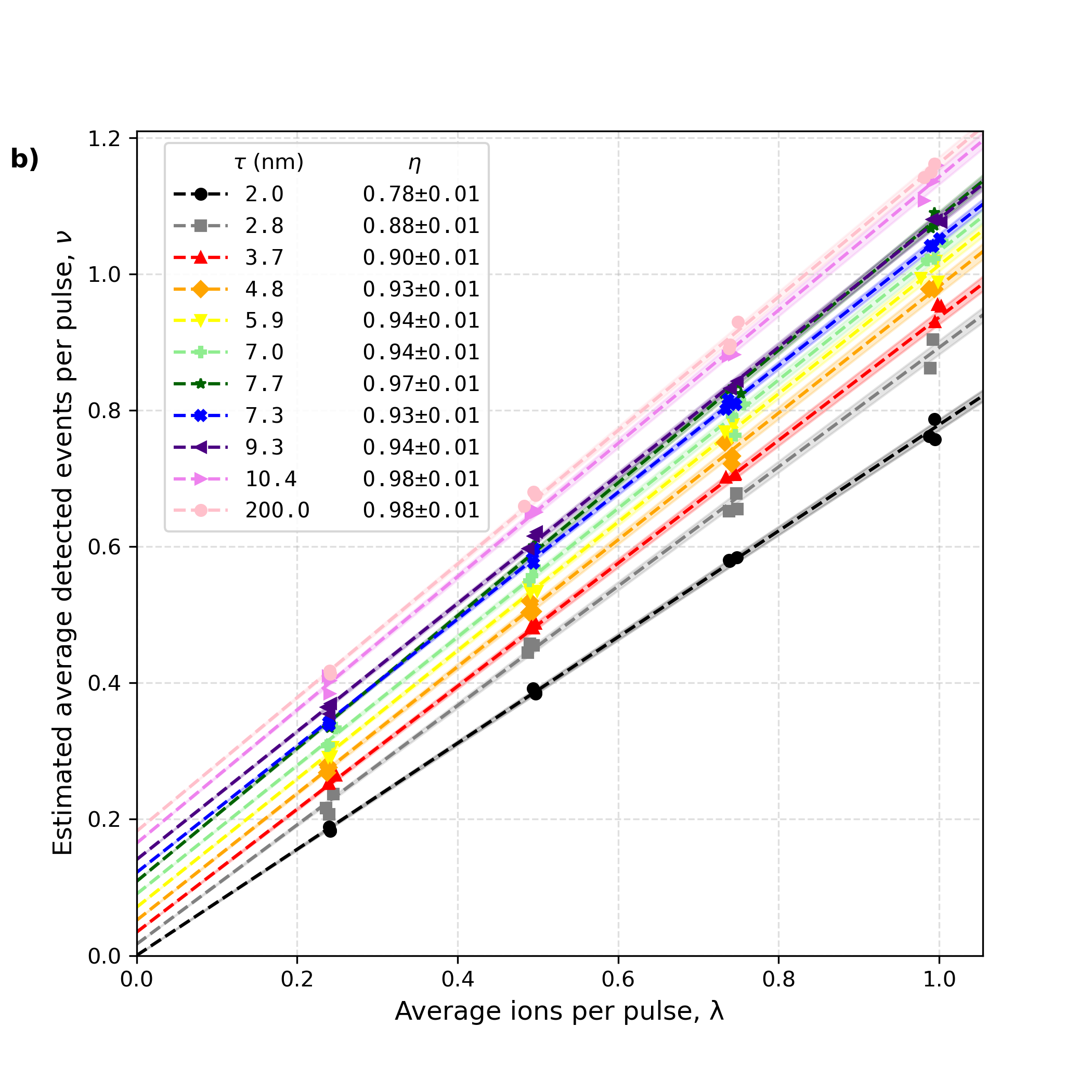}

    \caption{Secondary electron detection efficiency comparison for varying SiO${_2}$ capping-layer thicknesses (${\tau}$). (a) 25keV Sb$^+$. (b) 50keV Sb$^{2+}$.
    The plots show the relationship between the average ions per pulse (${\lambda}$) and the average number of detected events per pulse ($\nu={-\ln(p_0)}$, where $p_0$ is the fraction of empty pulses with no implant detections). 
    Each dashed line represents a linear regression fit to the data points for a specific sample, found as described in the text. The detection efficiency (${\eta}$) is given by the gradient of the line, and indicated in the legend. 
%    The regression was performed using a linear model (Equation~\ref{eq:YRegression}) to derive detection statistics.  
    For clarity of display, the intercept from each fit was subtracted off (both the fit and the data), and then a systematically increasing vertical offset was then added to each ${\tau}$ data set. 
    The shaded region around each fit line depicts the standard deviation in  the calculated gradient.}
\label{fig:DE_comparison}
\end{figure*}

For each sample, four  arrays of pixels were implanted with pulses having average doses of 0.25, 0.5, 0.75, and 1 ion per pulse. Pixels were spaced by 1 µm to avoid lateral overlap of implanted ions. During implantation, each pixel was repeatedly pulsed until an above-threshold secondary electron signal was detected by either of two channel electron multiplier (CEM) detectors positioned near the sample. The CEMs were electrostatically biased in a push–pull configuration to collect oppositely charged secondary particles. Once a pixel registered a hit, it was excluded from further pulsing, and the process continued until all pixels in the array produced a detectable event. The number of hits was therefore determined by the array size.
The total number of pulses required to make the whole array with one detection per pixel was recorded, and from this the overall fraction of empty pulses, $p_0$, was determined.

The uncertainty in $p_0$ scales with $1/\surd n$ where $n$ is the  total number of pulses used to make the array. To maintain a comparable uncertainty across the different dose conditions, array sizes of 40×40, 64×64, 70×70, and 80×80 pixels were used for mean doses of $\lambda=$0.25, 0.5, 0.75, and 1 ion per pulse, respectively. A higher $\lambda$ means fewer pulses per pixel, and so more pixels are needed.
With these choices and $\eta\sim1$ we expect 10,000 pulses per array, and the resulting error in $\nu\lesssim1\%$ in each case.

The beam current was regularly measured by diverting the beam into a Faraday cup connected to a Keithley Picoammeter. The beam was switched on and off for 10 s intervals, and the average change in the observed current provided the value of the incident ion flux. This calibration was performed before and after implantation of each set of four arrays, each run lasting approximately five minutes depending on array size. The typical current was 220fA and so 1 ion per pulse requires $t=$727ns. 

The results for $\nu(\lambda)$ are shown in Figure ~\ref{fig:DE_comparison}.

\section{Data Analysis}

\begin{figure*}[!b]
    \centering
    % Left image
    \includegraphics[width=0.48\textwidth]{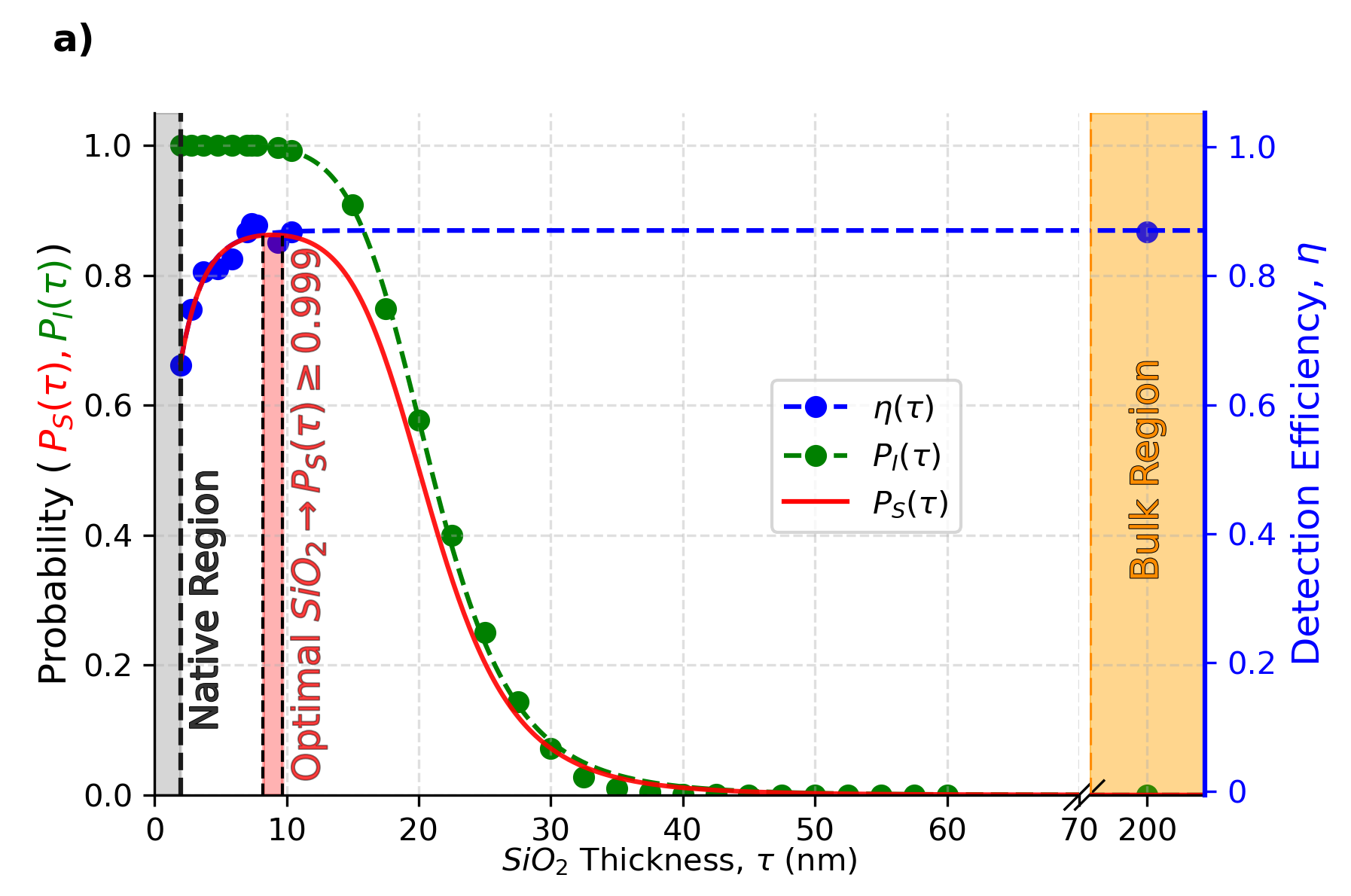}
    \hfill
    % Right image
    \includegraphics[width=0.48\textwidth]{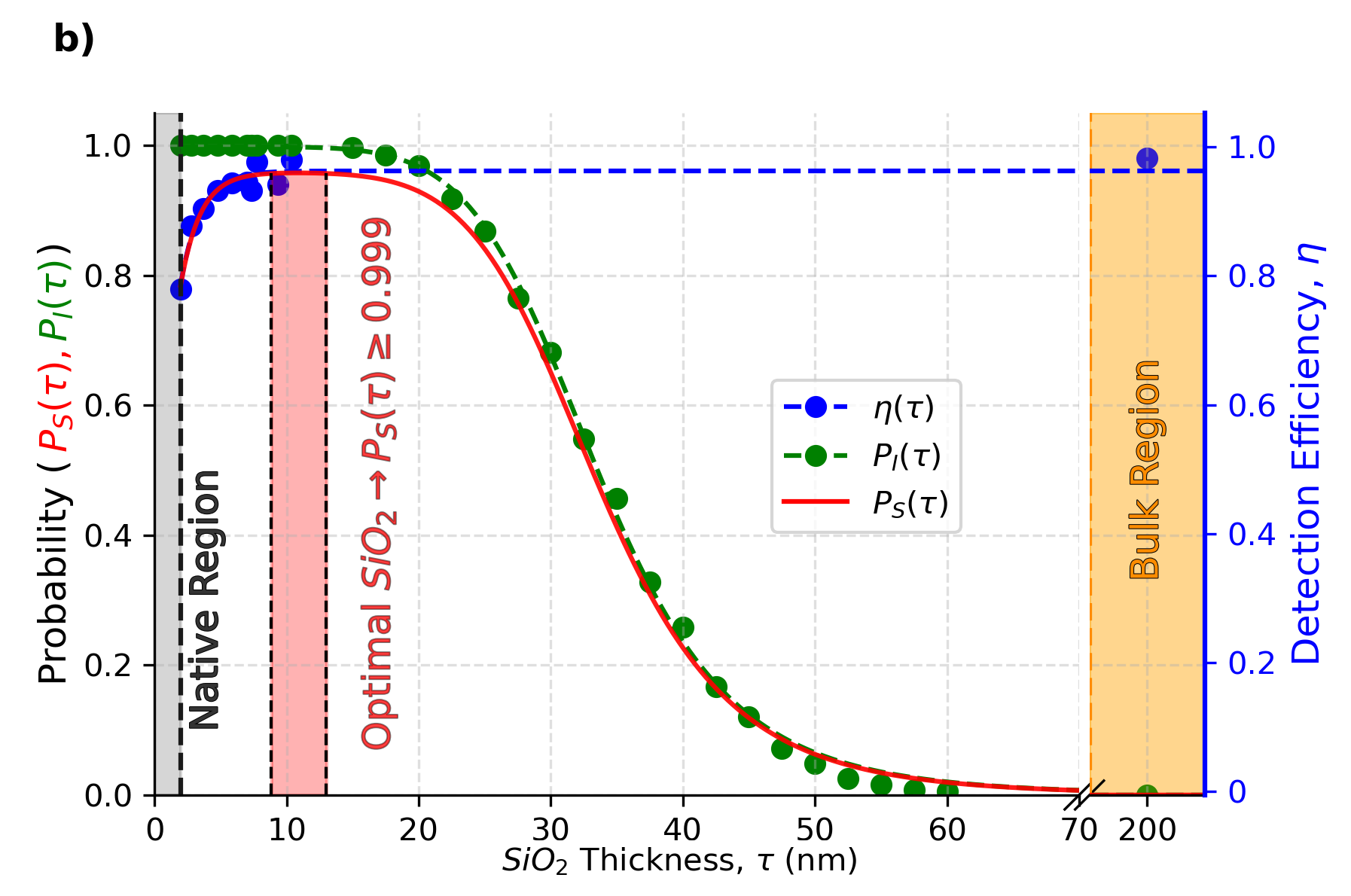}
    \caption{Optimal $\text{SiO}_2$ capping-layer thickness ($\tau$) determination for maximal implant success probability. 
    %The plots show three key quantities as a function of the capping-layer thickness, $\tau$. 
    The secondary electron detection efficiency, $\eta(\tau)$ (blue points, right ordinate axis), data from  Figure~\ref{fig:DE_comparison}. The error bars on these points are too small to see.
    The blue dashed line is a guide to the eye of form $A-a\exp(-\tau/\tau_0)$.
    The probability of an ion arriving into the Si substrate, $P_I(\tau)$ (green  points, left ordinate axis), was determined from SRIM simulations. 
    The green line is a guide to the eye of form $1 /(1 + (\tau/\tau_0)^k)$. 
    The total implant success probability, $P_S(\tau) $ (red line, left ordinate axis), from the produce of the green and blue lines (according to Eqn \ref{eq:prob_success}). This represents the probability of a  single  implant into the Si substrate and its simultaneous detection by the SE system. 
    The orange shaded region illustrates the  thickness range required for bulk SiO$_{2}$, for which all implanted atoms remain in the SiO$_2$, while the grey shaded region indicates the native oxide thickness range. 
    The red shaded area defines the range of $\tau$ that yields a total implant success probability, $P_S(\tau)$, within 99.9\% of its maximum value.}
\label{fig:Optimal_tau_determination}
\end{figure*}

\subsection*{Detection efficiency extraction}

The blanker that defines the ion pulses introduces a latency time $t_0$, corresponding to the transit time of ions through the blanker region. 
Only ions that enter after the blanker opens and exit before it closes reach the sample, effectively shortening the pulse duration by $t_0$. 
This produces a non-zero intercept in a graph of $\nu$ vs. $\lambda$ but has no effect on the slope.

Dark counts (false positives) can affect the results by increasing the apparent event rate to $N = K + \eta L$, where $K$ is the dark count rate. 
The true efficiency is then $\eta = \eta_a - K/L$, where $\eta_a$ is the apparent efficiency. 
%Clearly, if $K \ll L$ \ben{as in our case} then the dark counts can be ignored. 
The measured dark count rate ($>2$ s$^{-1}$) was several orders of magnitude smaller than the typical ion arrival rate ($>10^{6}$ s$^{-1}$), so $K/L < 10^{-2}$ and can be neglected.

We ensured that the number of pulses $n$ was sufficiently large that the statistical error in $p_0$ from the experiment described above was negligible compared to other uncertainties. 
In contrast, fluctuations in the ion current, $L$, were significant. 
During a single day of measurement, the standard deviation of the current measured between arrays was typically around 10\%, and even though the fluctuation between current calibrations in the experiment was much smaller, this was likely the dominant uncertainty in Eq.~\ref{eqn:p0def}. 
For this reason, to extract $\eta$ from the data, we performed a standard linear regression using
\begin{equation}
    T = -\frac{1}{L}\ln(p_0) 
    \label{eq:Tdef}
\end{equation}
as the dependent variable and the pulse duration $t$ as the independent variable.
This choice avoids having uncertainty in both axes, since fluctuations in the ion current $L$ dominate the experimental error. 
Physically, $T = \nu/L = \eta t$ can be interpreted as the \emph{effective active detection time} within each pulse: missing impacts due to undetected secondary electrons is equivalent to a detector with a fractional dead time of $1-\eta$ per pulse.
We assume the standard deviation in $T$ is constant (and proportional to that of $L$). 
According to Eq.~\ref{eqn:p0def}, the model to be fitted is
\begin{equation}
    T = \eta (t - t_0) = \eta t + c,
\end{equation}
where $\eta$ and $c $ are free parameters. 
In short, for presentation purposes we plot $\nu = L T$ versus $\lambda = L t$ in Figure \ref{fig:DE_comparison}, but the regression to obtain $\eta$ was performed on $T$ vs.\ $t$. 
As seen on Figure \ref{fig:DE_comparison}, the efficiency can be as much as 98\%.

The fit was done separately for each SiO$_2$ capping-layer thickness, $\tau$. 
% \oscar{Error on c in python not the same method as linear regression error calculation for $\hat{c}$, change code? } 
From the resulting values of the intercept $c(\tau)$ and slope $\eta(\tau)$ we obtained values of the blanker latency time $t_0(\tau)$. 
This time depends only on the ion energy and mass, not on the current or details of the sample host.  
For example, in the case of 25~keV Sb ions, 11 different samples yielded intercepts corresponding to a mean and standard deviation of $t_0 = 51 \pm 5$~ns. 
This standard deviation is significantly smaller than the rise time of the blanker pulse (10~ns), i.e. consistent with the hypothesis that $t_0$ is  constant. 
This measured transit time suggests an effective blanking region $\sim$10mm long. This is consistent with the blanker geometry in SIMPLE.

\begin{figure}
\includegraphics[width=\linewidth]{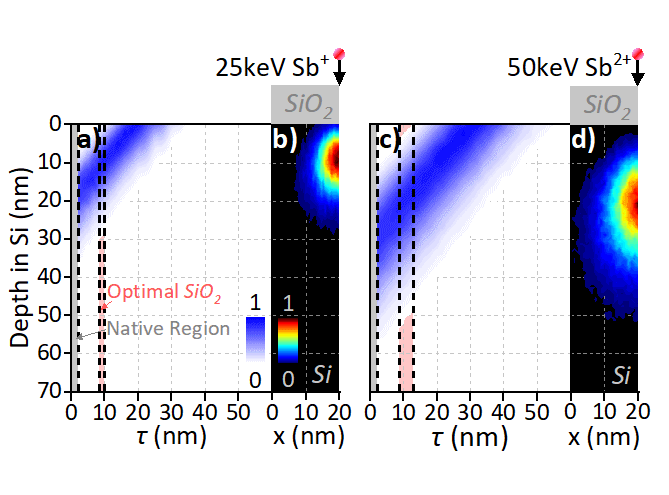}
\centering
\caption{TRIM simulations of Sb in the Si substrate after implantation at (a-b) 25keV and (c-d) 50keV through SiO$_2$. (a) and (c) show the implanted Sb depth profile as a function of SiO$_2$ thickness ($\tau$), with the colour intensity corresponding the normalised frequency density. (b) and (d) show 2D profiles of the normalised frequency density against lateral distance (x) and depth in Si for a 10nm SiO$_2$ thickness, which is at the optimum determined from Figure \ref{fig:Optimal_tau_determination}. The origin of the y axis corresponds where the Si interfaces with SiO$_2$.}
\label{fig:Sb_implant_depth_vs_SiO2_thickness}
\end{figure}

\subsection*{Successful Near-Single Ion Implants}

To determine the probability of a successful implant into the Si substrate through the SiO$_2$ capping layer, a distribution of detection efficiency as a function of $\tau$ is required. 
It is apparent from Figure ~\ref{fig:DE_comparison} that the slope $\eta$ increases with SiO$_2$ thickness, $\tau$. 
Figure~\ref{fig:Optimal_tau_determination} shows this trend explicitly. The error bars in experimental $\eta$  values, obtained from the regression on Figure \ref{fig:DE_comparison}, are too small to see.

Additionally, it is necessary to determine the probability that an ion of a given energy penetrates the capping layer, $P_{I}(\tau)$. 
This can be obtained by simulating ion stopping depths using Transport of Ions in Matter (TRIM) \cite{ZIEGLER2010}, which calculates the interactions of ions with amorphous targets using the Monte Carlo binary collision approximation.

In the TRIM simulations, each ion trajectory was tracked until it came to rest, producing a Monte Carlo distribution of stopping points in three dimensions. 
The depth of each ion below the surface was recorded and binned to form a one-dimensional histogram of stopping depths. 
For a given SiO$_2$ capping-layer thickness $\tau$, the probability $P_I$ that an implanted ion penetrates through the oxide and comes to rest in the Si substrate is obtained by counting the fraction of simulated ions that stop deeper than the SiO$_2$/Si interface.
Repeating this calculation for different oxide thicknesses yields the distribution $P_{I}(\tau)$.
The overall probability of a successful implant into the Si substrate is then
\begin{equation} \label{eq:prob_success}
P_{S}(\tau) = P_{I}(\tau)\,\eta(\tau),
\end{equation}
which allows the optimal capping-layer thickness $\tau$ to be determined from the maximum of $P_{S}(\tau)$.

In the TRIM simulations, targets consisted of a 10,000~\AA{} thick Si substrate layer ($\rho_{\rm Si} = 2.3212~\mathrm{g\,cm^{-3}}$) with an amorphous SiO$_{2}$ surface layer ($\rho_{\rm SiO_{2}} = 2.32~\mathrm{g\,cm^{-3}}$)~\cite{ZIEGLER2010}. 
Simulated results were obtained for incident ions of 25~keV Sb$^{+}$ and 50~keV Sb$^{2+}$. 
The SiO$_{2}$ thicknesses in the TRIM models replicated the experimental conditions, and $\tau$ was varied from 15~nm to 60~nm in 2.5~nm increments to generate a smooth $P_{I}(\tau)$ distribution. 
For each SiO$_2$ thickness, 50,000 implants were simulated.
The results for $P_I(\tau) $ and hence $P_S(\tau)$ are shown in Figure \ref{fig:Optimal_tau_determination}.
We find that the peak in the implant success probability is rather broad, which means that it is robust against poor deposition with uneven thickness.

TRIM was also used to calculate the depth of Sb in the Si substrate after implantation through the SiO$_{2}$ capping layer. The 25keV and 50keV Sb profiles were calculated using TRIM for each thickness of SiO$_{2}$ capping layer used in experiments. Then the depth profiles of the implanted Sb atoms in the Si substrate (without the SiO$_{2}$ capping layer) were calculated by subtracting the SiO$_{2}$ thicknesses from the TRIM ion ranges. The results are plotted as a function of SiO$_{2}$ thickness in Figure \ref{fig:Sb_implant_depth_vs_SiO2_thickness} (a) and (c) and show that the relationship between the depth of the Sb implant and the thickness of the SiO$_{2}$ capping layer  is close to linear. The reason for the simple nature of this relationship is that the distributions for the implant depths in both silicon and SiO$_2$ are very similar, so it matters little if the distribution straddles the interface. Figures \ref{fig:Sb_implant_depth_vs_SiO2_thickness} (b) and (d) show two-dimensional TRIM implant profiles of where the ion will stop in the Si substrate.

\subsection*{Other species}
\begin{figure}
\includegraphics[width=0.9\linewidth]{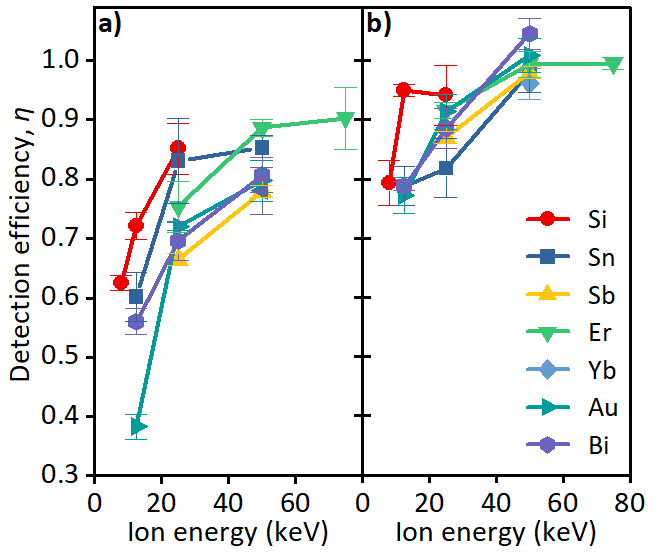}
\centering
\caption{Detection efficiency measurements as a function of ion energy for a Si sample with (a) native oxide only and (b) an $\text{SiO}_2$ target.  }
\label{fig:DE_vs_energy_various_species}
\end{figure}

SE detection from a wide range of implant species in both Si and SiO$_2$   has been reported by others\cite{Adshead2025}.  We also consider  other available implants here, in order to investigate the  generality of the improved SE detection using SiO$_2$ capping layers, and its  dependence on atomic mass.
Figure \ref{fig:DE_vs_energy_various_species} shows the SE efficiency for Si, Sb, Sn, Er, Yb, Au, Bi. The efficiency is shown as a function of ion energy: the accelerating voltage was either 8, 12.5 or 25 kV, and experiments were performed with available singly, doubly or triply charged ions emitted from the source, which were selected using a Wien filter.

\section{Discussion}

In the above experiment we quantify the  efficiency with which we detect SE, which clearly must increase with SE yield. Equally clearly, efficiency saturates at large yield which makes quantitative conclusions about yield challenging. Nevertheless, we can obtain useful  inferences from the trends with ion species, host and energy.  
Firstly, we observe that  the yield dependence on  SiO$_2$ thickness saturates when $\tau$ is just a few nm, whereas the distribution of ion stopping distances covers a few tens of nm. This immediately suggests that the emitted SE must be generated near the surface, so that the material where the majority of the ions and energy are deposited is irrelevant. This is consistent with Monte Carlo simulation studies which show that SE are generated less than about 10nm from the surface \cite{Ohya2004,Ullah2009}.

Because the yield depends only on the energy deposited near the surface, it should scale with the electronic stopping power, $S_{\mathrm{e}}$, rather than with the total ion energy $E$. 
According to Ullah \emph{et al.}~\cite{Ullah2009}, the kinetic component of the electron yield can be expressed as
\begin{equation}
    \gamma_{\mathrm{KE}} = \frac{P\,\ell_{\mathrm{e}}\,S_{\mathrm{e}}}{2J},
    \label{eq:yield}
\end{equation}
where $J$ is the mean energy required to produce a free electron in the solid, $\ell_{\mathrm{e}}$ is the electron mean free path, and $P$ is the mean escape probability for overcoming the surface barrier. 
Thus, $S_{\mathrm{e}}$ determines how much electronic energy is available for generating secondary electrons, while the product $P\ell_{\mathrm{e}}/J$ describes the material-dependent efficiency of converting that energy into emitted charge.

The stopping power $S_{\mathrm{e}}$ itself depends on the ion species, target composition, and velocity. 
In our experiments the ions are very slow---their speeds are well below the Thomas--Fermi velocity 
($v\!\ll\!v_\mathrm{TF}=\alpha c Z_1^{2/3}$, with $\alpha $ the fine structure constant and $c$ the speed of light). 
In this low-velocity regime the  Lindhard and Scharff (LS) relation applies~\cite{Lindhard1961,Sigmund2014}, giving
\begin{equation}
    S_{\mathrm{e}}\propto v\,\frac{Z_1Z_2}{(Z_1^{2/3}+Z_2^{2/3})^{3/2}},
    \label{eq:LS}
\end{equation}
where $Z_1$ and $Z_2$ are the atomic numbers of the projectile and target, respectively. 
Our experimental results show enhanced SE detection efficiency for faster ions of a given species, consistent with this prediction. 
Using $v=\sqrt{2E_1/A_1u}$, where $E_1$ and $A_1$ are the projectile energy and mass and $u$ is the atomic mass unit,  Eqn \ref{eq:LS} predicts  a small decrease (16\%) in stopping power from Si to Bi at fixed energy.
The modest decline in detection efficiency with increasing atomic number observed in Figure \ref{fig:DE_vs_energy_various_species} is therefore consistent with the LS prediction. A similar slightly decreasing trend of the detection efficiency with increasing atomic mass appears to have been observed elsewhere \cite{Adshead2025}.
We note that alternative treatments of stopping of slow ions with $v\ll v_\text{TF}$ produce~\cite{Sigmund2014} $S_{\mathrm{e}}\propto v Z_1^2/Z_2$ which is a stronger dependence on species, predicting  an order of magnitude  increase  in yield from Si to Bi, contrary to our observation.

The LS formula predicts a slight decrease in stopping power of SiO$_2$ compared with Si due to the lower average atomic number $Z$, which is contrary to the observations of Figures  \ref{fig:DE_comparison} and \ref{fig:Optimal_tau_determination} for the overall yield. At the same time the large band gap ($E_\mathrm{g}\approx 9~\mathrm{eV}$) of SiO$_2$ compared with Si ($E_\mathrm{g}\approx 1~\mathrm{eV}$) suppresses the generation of low-energy electrons by increasing $J$ which also tends to reduce the yield.  It therefore must be the case that there is a more important increase in $P\ell_{\mathrm{e}}$ (lower barrier and/or larger IMFP) in SiO$_2$.

Ohya and Ishitani~\cite{Ohya2004} simulated SE emission from SiO$_2$ and Si under bombardment by focused Ga$^+$ ions at 30keV, and found that the SE yield of SiO$_2$ was \emph{lower} than that of Si. 
In their model the  wide band gap affects $\ell_{\mathrm{e}}$: the lower density of available states in the conduction band reduces the probability of further inelastic scattering by impact ionization, thereby \emph{increasing} it. 
This longer IMFP means that electrons which do exceed the excitation threshold can propagate further without energy loss, increasing their chance of escaping if they are generated within a few nanometres of the surface. However, in their model, the increase in $\ell_{\mathrm{e}} $ was not sufficient to compensate the increased $J$ and increase the yield for SiO$_2$ relative to Si.

In a  more recent study, Ullah \emph{et al.}~\cite{Ullah2009} investigated SE emission due to implantation of Ne$^+$ into SiO$_2$ for rare-gas ion impacts in the 1--10~keV range and found that the yield was \emph{higher} for SiO$_2$ than for earlier data for Si \cite{Wittmaack1991}. 
They explained this opposite conclusion in terms of different assumptions about the escape  barrier $P$ and the energy dependence of the IMFP, $\ell_{\mathrm{e}}$. 
Clearly, our observation of a higher yield from SiO$_2$ support the conclusion that $P\ell_{\mathrm{e}}$ is significantly increased in order to compensate from the reduction in $S_{\mathrm{e}}$ and increase in $J$.

Ullah \emph{et al.} also predicted that, at fixed ion energy of 10keV, the SE yield from SiO$_2$ increases strongly with projectile mass, following the sequence   Ne$^+<$  Ar$^+<$  Kr$^+<$  Xe$^+$. 
This is not consistent with our observation of a slight drop in efficiency as we go from Si to Bi, as mentioned above.

The contrast between our results and the Ga$^+$ data of Ohya and Ishitani~\cite{Ohya2004}, as well as the rare gas study of Ullah \emph{et al.}~\cite{Ullah2009}, is  revealing. 
All three cases concern ions of overlapping mass range mass and energy, yet the two modelling studies reach opposite conclusions regarding the relative SE yield of Si and SiO$_2$, and the species (mass) dependence. 
This discrepancy appears to be due to differences in the underlying assumptions governing $S_{\mathrm{e}}$, $\ell_{\mathrm{e}}$, $J$, and the escape probability $P$. 
In both modelling studies, the electronic stopping and transport processes were treated using semi-empirical models whose parameters were optimised for specific materials. 
Our experimental measurements constrain  these parameters.

Beyond providing insight into these mechanisms, our findings also suggest a practical route for optimising single-ion detection. 
A thin oxide layer such as SiO$_2$ can act as an \emph{emission-enhancing coating} that improves SE collection efficiency without affecting implantation accuracy. 
Since such oxides can be readily removed by standard chemical etching (e.g.\ dilute HF), this approach could be extended to a wide variety of target substrates---including semiconductors, metals, and dielectrics---enabling deterministic ion implantation with high detection fidelity regardless of the host material.

\section{Conclusion}
We have demonstrated a robust, high-efficiency, and non-destructive approach for detecting single-ion implantation events in silicon using secondary electron emission within a focused ion beam system. By introducing a thin SiO$_2$ capping layer, we achieved detection efficiencies as high as 98\%, produced through calibrated ion-current measurements. The enhanced secondary-electron yield is attributed to an increased inelastic mean free path and escape probability in the oxide. TRIM simulations reveal that optimal detection coincides with oxide thicknesses that still permit near-unity implantation probability into the underlying silicon. The thickness of the oxide can also have considerable error of $\pm1$nm for 25keV implants and $\pm2$ for 50keV, and still be very close to optimum, i.e. it is robust against deposition error.  The method attains nanometre spatial precision without requiring electrical contacts or pre-fabricated device structures, and anticipate that for most applications the SiO$_2$ cap could later be removed chemically, as desired. Our findings establish secondary-electron-based detection as a scalable route for deterministic dopant placement and provide a general framework for extending single-ion detection to a broad class of materials and species relevant to quantum device fabrication.

\bibliography{bibliography.bib}

\providecommand{\latin}[1]{#1}
\makeatletter
\providecommand{\doi}
  {\begingroup\let\do\@makeother\dospecials
  \catcode`\{=1 \catcode`\}=2 \doi@aux}
\providecommand{\doi@aux}[1]{\endgroup\texttt{#1}}
\makeatother
\providecommand*\mcitethebibliography{\thebibliography}
\csname @ifundefined\endcsname{endmcitethebibliography}  {\let\endmcitethebibliography\endthebibliography}{}
\begin{mcitethebibliography}{19}
\providecommand*\natexlab[1]{#1}
\providecommand*\mciteSetBstSublistMode[1]{}
\providecommand*\mciteSetBstMaxWidthForm[2]{}
\providecommand*\mciteBstWouldAddEndPuncttrue
  {\def\EndOfBibitem{\unskip.}}
\providecommand*\mciteBstWouldAddEndPunctfalse
  {\let\EndOfBibitem\relax}
\providecommand*\mciteSetBstMidEndSepPunct[3]{}
\providecommand*\mciteSetBstSublistLabelBeginEnd[3]{}
\providecommand*\EndOfBibitem{}
\mciteSetBstSublistMode{f}
\mciteSetBstMaxWidthForm{subitem}{(\alph{mcitesubitemcount})}
\mciteSetBstSublistLabelBeginEnd
  {\mcitemaxwidthsubitemform\space}
  {\relax}
  {\relax}

\bibitem[Schofield \latin{et~al.}(2025)Schofield, Fisher, Ginossar, Lyding, Silver, Fei, Namboodiri, Wyrick, Masteghin, Cox, Murdin, Clowes, Keizer, Simmons, Stemp, Morello, Voisin, Rogge, Wolkow, Livadaru, Pitters, Stock, Curson, Butera, Pavlova, Jakob, Spemann, Räcke, Schmidt-Kaler, Jamieson, Pratiush, Duscher, Kalinin, Kazazis, Constantinou, Aeppli, Ekinci, Owen, Fowler, Moheimani, Randall, Misra, Ivie, Allemang, Anderson, Bussmann, Campbell, Gao, Lu, and Schmucker]{Schofield2025}
Schofield,~S.~R. \latin{et~al.}  \emph{Nano Futures} \textbf{2025}, \emph{9}, 012001\relax
\mciteBstWouldAddEndPuncttrue
\mciteSetBstMidEndSepPunct{\mcitedefaultmidpunct}
{\mcitedefaultendpunct}{\mcitedefaultseppunct}\relax
\EndOfBibitem
\bibitem[Morello \latin{et~al.}(2020)Morello, Pla, Bertet, and Jamieson]{Morello2020}
Morello,~A.; Pla,~J.~J.; Bertet,~P.; Jamieson,~D.~N. \emph{Advanced Quantum Technologies} \textbf{2020}, \emph{3}\relax
\mciteBstWouldAddEndPuncttrue
\mciteSetBstMidEndSepPunct{\mcitedefaultmidpunct}
{\mcitedefaultendpunct}{\mcitedefaultseppunct}\relax
\EndOfBibitem
\bibitem[Chang \latin{et~al.}(2024)Chang, Holzman, Lim, Holmes, Johnson, Jamieson, and Stern]{chang2024}
Chang,~T.; Holzman,~I.; Lim,~S.~Q.; Holmes,~D.; Johnson,~B.~C.; Jamieson,~D.~N.; Stern,~M. Strong coupling of a superconducting flux qubit to single bismuth donors. 2024\relax
\mciteBstWouldAddEndPuncttrue
\mciteSetBstMidEndSepPunct{\mcitedefaultmidpunct}
{\mcitedefaultendpunct}{\mcitedefaultseppunct}\relax
\EndOfBibitem
\bibitem[Yu \latin{et~al.}(2025)Yu, Wilhelm, Holmes, Vaartjes, Schwienbacher, Nurizzo, Kringhøj, van Blankenstein, Jakob, Gupta, Hudson, Itoh, Murray, Blume-Kohout, Ladd, Anand, Dzurak, Sanders, Jamieson, and Morello]{Yu2025}
Yu,~X. \latin{et~al.}  \emph{Nature Physics} \textbf{2025}, \emph{21}, 362--367\relax
\mciteBstWouldAddEndPuncttrue
\mciteSetBstMidEndSepPunct{\mcitedefaultmidpunct}
{\mcitedefaultendpunct}{\mcitedefaultseppunct}\relax
\EndOfBibitem
\bibitem[Jakob \latin{et~al.}(2022)Jakob, Robson, Schmitt, Mourik, Posselt, Spemann, Johnson, Firgau, Mayes, McCallum, Morello, and Jamieson]{Jakob2022}
Jakob,~A.~M.; Robson,~S.~G.; Schmitt,~V.; Mourik,~V.; Posselt,~M.; Spemann,~D.; Johnson,~B.~C.; Firgau,~H.~R.; Mayes,~E.; McCallum,~J.~C.; Morello,~A.; Jamieson,~D.~N. \emph{Advanced Materials} \textbf{2022}, \emph{34}\relax
\mciteBstWouldAddEndPuncttrue
\mciteSetBstMidEndSepPunct{\mcitedefaultmidpunct}
{\mcitedefaultendpunct}{\mcitedefaultseppunct}\relax
\EndOfBibitem
\bibitem[Titze \latin{et~al.}(2022)Titze, Byeon, Flores, Henshaw, Harris, Mounce, and Bielejec]{Titze2022}
Titze,~M.; Byeon,~H.; Flores,~A.; Henshaw,~J.; Harris,~C.~T.; Mounce,~A.~M.; Bielejec,~E.~S. \emph{Nano Letters} \textbf{2022}, \emph{22}, 3212--3218\relax
\mciteBstWouldAddEndPuncttrue
\mciteSetBstMidEndSepPunct{\mcitedefaultmidpunct}
{\mcitedefaultendpunct}{\mcitedefaultseppunct}\relax
\EndOfBibitem
\bibitem[Cassidy \latin{et~al.}(2021)Cassidy, Blenkinsopp, Brown, Curry, Murdin, Webb, and Cox]{Cassidy2021}
Cassidy,~N.; Blenkinsopp,~P.; Brown,~I.; Curry,~R.~J.; Murdin,~B.~N.; Webb,~R.; Cox,~D. \emph{physica status solidi (a)} \textbf{2021}, \emph{218}\relax
\mciteBstWouldAddEndPuncttrue
\mciteSetBstMidEndSepPunct{\mcitedefaultmidpunct}
{\mcitedefaultendpunct}{\mcitedefaultseppunct}\relax
\EndOfBibitem
\bibitem[Murdin \latin{et~al.}(2021)Murdin, Cassidy, Cox, Webb, and Curry]{Murdin2021}
Murdin,~B.~N.; Cassidy,~N.; Cox,~D.; Webb,~R.; Curry,~R.~J. \emph{Physica Status Solidi (B) Basic Research} \textbf{2021}, \emph{258}\relax
\mciteBstWouldAddEndPuncttrue
\mciteSetBstMidEndSepPunct{\mcitedefaultmidpunct}
{\mcitedefaultendpunct}{\mcitedefaultseppunct}\relax
\EndOfBibitem
\bibitem[Adshead \latin{et~al.}(2025)Adshead, Wan, Coke, and Curry]{Adshead2025}
Adshead,~M.; Wan,~L.~K.; Coke,~M.; Curry,~R.~J. \emph{arXiv:2510.01035} \textbf{2025}, \relax
\mciteBstWouldAddEndPunctfalse
\mciteSetBstMidEndSepPunct{\mcitedefaultmidpunct}
{}{\mcitedefaultseppunct}\relax
\EndOfBibitem
\bibitem[Sahin \latin{et~al.}(2017)Sahin, Geppert, Müllers, and Ott]{Sahin2017}
Sahin,~C.; Geppert,~P.; Müllers,~A.; Ott,~H. \emph{New Journal of Physics} \textbf{2017}, \emph{19}, 123005\relax
\mciteBstWouldAddEndPuncttrue
\mciteSetBstMidEndSepPunct{\mcitedefaultmidpunct}
{\mcitedefaultendpunct}{\mcitedefaultseppunct}\relax
\EndOfBibitem
\bibitem[Räcke \latin{et~al.}(2022)Räcke, Meijer, and Spemann]{Racke2022}
Räcke,~P.; Meijer,~J.; Spemann,~D. \emph{Journal of Applied Physics} \textbf{2022}, \emph{131}\relax
\mciteBstWouldAddEndPuncttrue
\mciteSetBstMidEndSepPunct{\mcitedefaultmidpunct}
{\mcitedefaultendpunct}{\mcitedefaultseppunct}\relax
\EndOfBibitem
\bibitem[Stopp \latin{et~al.}(2022)Stopp, Lehec, and Schmidt-Kaler]{Stopp2022}
Stopp,~F.; Lehec,~H.; Schmidt-Kaler,~F. \emph{Quantum Science and Technology} \textbf{2022}, \emph{7}, 034002\relax
\mciteBstWouldAddEndPuncttrue
\mciteSetBstMidEndSepPunct{\mcitedefaultmidpunct}
{\mcitedefaultendpunct}{\mcitedefaultseppunct}\relax
\EndOfBibitem
\bibitem[Ziegler \latin{et~al.}(2010)Ziegler, Ziegler, and Biersack]{ZIEGLER2010}
Ziegler,~J.~F.; Ziegler,~M.; Biersack,~J. \emph{Nuclear Instruments and Methods in Physics Research Section B: Beam Interactions with Materials and Atoms} \textbf{2010}, \emph{268}, 1818--1823, 19th International Conference on Ion Beam Analysis\relax
\mciteBstWouldAddEndPuncttrue
\mciteSetBstMidEndSepPunct{\mcitedefaultmidpunct}
{\mcitedefaultendpunct}{\mcitedefaultseppunct}\relax
\EndOfBibitem
\bibitem[Ohya and Ishitani(2004)Ohya, and Ishitani]{Ohya2004}
Ohya,~K.; Ishitani,~T. \emph{Applied Surface Science} \textbf{2004}, \emph{237}, 602--606\relax
\mciteBstWouldAddEndPuncttrue
\mciteSetBstMidEndSepPunct{\mcitedefaultmidpunct}
{\mcitedefaultendpunct}{\mcitedefaultseppunct}\relax
\EndOfBibitem
\bibitem[Ullah \latin{et~al.}(2009)Ullah, Dogar, and Qayyum]{Ullah2009}
Ullah,~S.; Dogar,~A.; Qayyum,~A. \emph{Nuclear Instruments and Methods in Physics Research Section B: Beam Interactions with Materials and Atoms} \textbf{2009}, \emph{267}, 3059--3062\relax
\mciteBstWouldAddEndPuncttrue
\mciteSetBstMidEndSepPunct{\mcitedefaultmidpunct}
{\mcitedefaultendpunct}{\mcitedefaultseppunct}\relax
\EndOfBibitem
\bibitem[Lindhard and Scharff(1961)Lindhard, and Scharff]{Lindhard1961}
Lindhard,~J.; Scharff,~M. \emph{Physical Review} \textbf{1961}, \emph{124}, 128--130\relax
\mciteBstWouldAddEndPuncttrue
\mciteSetBstMidEndSepPunct{\mcitedefaultmidpunct}
{\mcitedefaultendpunct}{\mcitedefaultseppunct}\relax
\EndOfBibitem
\bibitem[Sigmund(2014)]{Sigmund2014}
Sigmund,~P. \emph{Particle Penetration and Radiation Effects Volume 2}; Springer Series in Solid-State Sciences, 2014; Vol. 179\relax
\mciteBstWouldAddEndPuncttrue
\mciteSetBstMidEndSepPunct{\mcitedefaultmidpunct}
{\mcitedefaultendpunct}{\mcitedefaultseppunct}\relax
\EndOfBibitem
\bibitem[Wittmaack(1991)]{Wittmaack1991}
Wittmaack,~K. \emph{Nuclear Instruments and Methods in Physics Research Section B: Beam Interactions with Materials and Atoms} \textbf{1991}, \emph{58}, 317--321\relax
\mciteBstWouldAddEndPuncttrue
\mciteSetBstMidEndSepPunct{\mcitedefaultmidpunct}
{\mcitedefaultendpunct}{\mcitedefaultseppunct}\relax
\EndOfBibitem
\end{mcitethebibliography}

\end{document}